\documentclass{aastex631}
\usepackage{hyperref}
\usepackage{url}
\usepackage{xcolor}

\begin{document}

\title{Reinforcement Learning for Data-Driven Workflows in Radio Interferometry. I. \\ Principal Demonstration in Calibration}

\author{Brian M. Kirk}
\affiliation{New Mexico Institute of Mining and Technology} 
\affiliation{National Radio Astronomy Observatory}
\affiliation{Grote Reber Fellow}

\author{Urvashi Rau}
\affiliation{National Radio Astronomy Observatory}

\author{Ramyaa Ramyaa}
\affiliation{New Mexico Institute of Mining and Technology}


\begin{abstract}
Radio interferometry is an observational technique used to study astrophysical phenomena. Data gathered by an interferometer requires substantial processing before astronomers can extract the scientific information from it. Data processing consists of a sequence of calibration and analysis procedures where choices must be made about the sequence of procedures as well as the specific configuration of the procedure itself. These choices are typically based on a combination of measurable data characteristics, an understanding of the instrument itself, an appreciation of the trade-offs between compute cost and accuracy, and a learned understanding of what is considered “best practice”. A metric of absolute correctness is not always available and validity is often subject to human judgment. The underlying principles and software configurations to discern a reasonable workflow for a given dataset is the subject of training workshops for students and scientists. Our goal is to use objective metrics that quantify best practice, and numerically map out the decision space with respect to our metrics. With these objective metrics we demonstrate an automated, data-driven, decision system that is capable of sequencing the optimal action(s) for processing interferometric data. This paper introduces a simplified description of the principles behind interferometry and the procedures required for data processing. We highlight the issues with current automation approaches and propose our ideas for solving these bottlenecks. A prototype is demonstrated and the results are discussed.
\end{abstract}


\section{Introduction} \label{sec:p1}


\subsection{Principles behind interferometric images} \label{sec:p1.1} Radio astronomy is the study of astronomical objects at radio frequencies and the physical processes generating that emission. A radio interferometer uses a collection of antennas to measure the Fourier space of the emission distribution across the sky. The fundamental observation of every pair of radio antennas is a measurement of the spatial coherence of the electromagnetic field at their respective locations. Signals from both antennas in each pair are combined to form a visibility, which is a 2D Fourier component of the sky’s brightness distribution. By accumulating many different Fourier components through a collection of antennas with different orientations and separations, the Fourier space representation of the sky can be observed. The (inverse) Fourier transform may then be applied to give astronomers the spatial distribution of the emission intensity in the sky, i.e an image. However, an infinite number of Fourier components are needed to get a perfect representation of the sky. Radio interferometers have a finite number of antennas and therefore can only measure a finite number of Fourier components. The missing information introduces ambiguity into the resulting images. The process of image reconstruction is an inverse model fitting process, subject to biases introduced by the algorithms and optimization strategies. Additionally, before imaging, the calibration of each antenna is vitally important. The effect of the antenna and receiver system must be removed from the measured Fourier components before image reconstruction is attempted. Data that are corrupted by radio frequency interference or other errors must also be identified and removed as part of the data processing. There are a multitude of different ways one can process the data, for example using different calibration strategies, choices between outlier detection methods, and selecting different imaging algorithms for the inverse modeling. Choices of what to do are often based on imperfect information and an astronomer’s best judgment, and this can lead to processing the information in multiple ways that may be considered equally valid (or invalid).

\subsection{Stages of data processing} \label{sec:p1.2} The main stages of data processing consist of flagging erroneous data, calibrating the instrument hardware, and reconstructing the image through deconvolution. These stages may be repeated any number of times and in different sequences for any given dataset. 

In the flagging stage, data are identified to be eliminated or ignored from subsequent processing. This can be due to corruption by radio frequency interference (RFI) or other related errors that can lead to outliers. There are different customizable algorithms suited to identifying different types of RFI. Depending on the situation, data flagging can be performed multiple times on the raw data before calibration, intermediate calibration products, and or the final calibrated data. Alternatively, the outlier detection and flagging stage is not always necessary, and depends on the presence or absence of RFI. 

In the calibration stage, observations from a known reference source are used to derive instrumental antenna gains (complex multipliers on the ideal voltage streams, which vary with time, frequency and antenna) and correct for this instrumental effect from the data. Corrections for direction dependent calibration terms are typically expressed as non-linear matrix equations and can be solved with non-linear least squares approaches in most cases. The accuracy of these corrections depends on the signal-to-noise ratio (SNR) of the observed calibrator data presented to the solver. With observations recording data across multiple time-steps and across a range of frequencies, the time and frequency data may be averaged to increase SNR for a more robust correction. However, excessive averaging in time or frequency can misrepresent the very antenna  gain distortions that require modeling and correction. Therefore, a careful data-dependent choice is required in the amount of averaging that is applied, to maximize SNR for the solver but without destroying the gain distortions themselves. Additionally, calibration operations may be iterative, to model and remove different instrumental effects along the signal acquisition chain. 

Finally, the image reconstruction stage is an iterative model fitting process. Transforms between the Fourier components and the image domain are interleaved with sky modeling steps in a gradient descent approach for chi-square minimization. There are multiple reconstruction algorithms one can choose from in this stage. Typically, algorithm selection depends on the type of image being made (such as one image containing the average intensity across all frequencies, or separate images for each frequency channel) and the type of structure being reconstructed (point-like sources vs diffuse extended emission or a combination of both). These algorithms differ in image reconstruction quality, computational runtime, and numerical convergence efficiency. It is a subjective process to choose the algorithms and the optimal parameters that will give the ‘best’ possible image for a given data set. 

\begin{figure}[hbt!]
    \centering
    \includegraphics[width=0.65\linewidth]{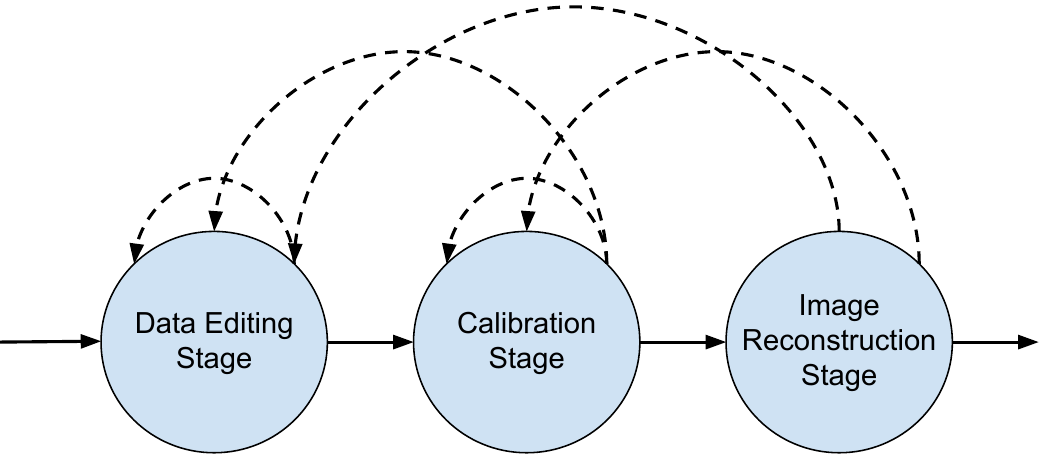}
    \caption{Highlighting the possibility of loops in processing steps (dotted lines) rather than a linear flow (solid lines).}
    \label{fig:stage-loops}
\end{figure}

The choices made throughout all of the stages is commonly referred to as a ‘\textit{recipe}’. Any single end-to-end data processing recipe may be thought of as one possible path out of all possible configurations. For end-to-end processing, there is a predictable sequence of stages, where each stage may contain several procedures depending on the characteristics of the data. If certain data conditions exist, some procedures, or entire stages, may be revisited as subsequent procedures, thus forming processing loops (Fig.\ref{fig:stage-loops}). 

The full configuration space of the available software tools is quite vast, leading to many possible processing paths. The understanding needed to create a recipe to process the data is the subject of training workshops where students and scientists are taught the principles of each stage, how to decide and configure the multiple procedures within a stage, and how to sequence the stages to prepare data for scientific analysis. It is this decision space that we are grounding on a numerical basis to enable automated data-driven processing. This work was inspired by the field of Reinforcement Learning (RL) and specifically AlphaGo \citep{44806}. In this paper we demonstrate on a constrained scenario while only considering the calibration and RFI flagging stages where our objective is to remove any effect on the visibility data by antenna gains and outliers. This is to validate whether RL methods, in general, are suitable for our domain and we begin this validation with the simplest RL method, Q-learning. This validation is fundamental before moving on to more sophisticated RL methods, which are derived from Q-learning.

Q-learning is a Reinforcement Learning method that learns the optimal action-value function from experience. By repeatedly sampling a Q-table's states, different actions, and action-values for a variety of transitions, the action value for a given state-action pair is updated based on the downstream effect it has on the final result. This is a way for the combined effects of actions that lead to good outcomes (or bad) to be reflected in the individual actions that led to that outcome. By iteratively updating the action values, the positive (or negative) numerical feedback can be used to steer decisions towards the desired outcome. This is a hallmark of Reinforcement Learning.

\subsection{Current automation and its limitations} \label{sec:p1.3}
Current automated approaches are little more than hard-coding a “few” of the most-used non-looping recipes for processing, with a small set of heuristics-based decision points. In the field these hard-coded recipes are called ‘\textit{pipelines}’. Pipelines are designed for the general use-case rather than for being specific. A pipeline includes all processing steps and parameter settings the average dataset would need, regardless of a \textit{particular} dataset's needs. For example, a dataset with little-to-no RFI may not need any flagging at all. Such hard-coded recipes are inflexible to different characteristics of an observation which may warrant different processing actions or parameter tuning.

The pipelines (ALMA \citep{2016adass..26P1.14G}, VLA \citep{2018AAS...23134214K}, SRDP \citep{2021AAS...23753510T}, IDIA \citep{9560276}, HiFAST\citep{2024SCPMA..6759514J}, uGMRT’s CAPTURE\citep{2021ExA....51...95K}, Artip\citep{2018ascl.soft02004S}, etc.) are designed to be robust for datasets that fall within a narrow range of data characteristics and observation modes. For example, the ALMA pipeline is designed to address a specific range of use-cases and processes 95\% of \textit{qualifying} datasets through the pipeline, with the remaining 5\% processed manually. Out of the qualifying 95\% that go through the pipeline, about a third of datasets typically require human intervention to improve the end result \citep{nrao}. The VLA produces a more diverse set of input datasets (with more RFI), all of which are passed through its standard pipeline \citep{2018AAS...23134214K} and inspected manually. Out of these, a quarter are reprocessed by humans adjusting the pipeline scripts, and the rest are handed to the end user as-is to reprocess themselves, if needed \citep{nrao}. Finally, both instruments have begun to offer a reprocessing service where end users can request further adjustments in addition to what was delivered to them (\href{https://science.nrao.edu/srdp/features}{AUDI}), implying that there is often a need for further customization per dataset. In short, the one-size-fits-all processing via pipelines leaves much room for improvement. This is indicated by pipelines having yet to achieve outputs that require \textit{no further human intervention} to customize processing before being used for astrophysical research. A similar problem is faced in astronomy at other wavelengths \citep{Borgman2021From}, \citep{Freudling_2024}.

Failing to achieve true science-ready automated processing, while that is an issue in itself, causes an additional penalty: a pipeline’s results may need to be reviewed and reprocessed by a observatory staff. For example, at NRAO run facilities, trained analysts are used to review the pipeline results. They inspect the results for errors, identify optimizations that can be made in individual steps or entire sequences, and repeat data processing \textit{after} the pipeline has already processed the data. This is because the pipeline results may not meet quality thresholds. This reprocessing increases operating costs. Furthermore, even after reprocessing is done by a trained analyst, additional enhancement may still be done by the end-user.

While one can estimate some aspects of an observation, we do not know for certain the signal distribution, observing conditions, antenna performance, RFI contamination, sources present, and so on which all alter how the data should be optimally processed. Therefore, what is needed is an automated system sensitive to the contents of the observation that can decide what the best processing step should be. In this paper we explore Reinforcement Learning methods as a data-driven approach to automation that can potentially handle observations containing a wide variety of characteristics and dynamically determine what the best course of action is.

We propose this can be done by measuring the performance of an action in a given situation. By setting an objective and measuring the effects a particular action has on a wide range of datasets, a function is discovered for how well that action performs given the state of the data. This can be thought of as a processing action’s “return on investment” function, or action-value. By discovering this function for each of the different actions, across a large range of datasets (states), we learn what the optimal action would be for any given dataset. With each action's individual action-value function found, the basis is provided to evaluate a series of actions which is a recipe or workflow. With individual action-value functions established, we can then investigate how different sequences of actions (different workflows) will affect a given dataset. By gaining an understanding of when, or when not, to apply an action results in a flexible system that can choose the ideal processing step based on the data rather than following a hard-coded pipeline. 

Another benefit of Reinforcement Learning is that evaluating actions and estimating action-value functions can be automated. As more data becomes available these action-value functions and processing workflows can be refined, which ultimately shape the decision process. This same concept of iterative improvement is currently used by development teams (more data → refine heuristics) but on a much slower time scale. For example, the NRAO pipelines have taken 10+ years of manual heuristic development \citep{2023PASP..135g4501H}.

We speculate that a flexible data-driven approach, such as RL, may provide a way to get science-ready data products \textit{and} large scale automation. The following section introduces a fundamental demonstration with a particular decision in the calibration stage routinely used in processing. We illustrate a way to find the optimal processing action based on the data itself rather than following hard-coded directions. This is extended to finding the optimal processing workflow for the  sequence of multiple actions. The scope of our proof-of concept demonstration is limited to a highly simplified analysis scenario, designed to allow unambiguous verification of our algorithms while also applying all the core machine learning concepts we will need for a more complex data analysis sequence.

Other work involving Reinforcement Learning and Radio astronomy calibration is done in \citep{10.1093/mnras/stab1401}. Here, the authors are using different reinforcement learning methods to determine a regularization hyper-parameter which controls the smoothness of their correction models for calibration. In their application the agent is provided with time series data to learn from and make recommendations for (continuous) regularization factor(s) for the remaining time series of data. In addition to using different reinforcement learning methods, the most noticeable difference is \citep{10.1093/mnras/stab1401} use Reinforcement Learning to learn a hyper-parameter that is \textit{given to a pipeline}. In contrast, our work is focused on assembling the ideal pipeline for any given dataset, not tuning parameters that are given to a pipeline.

In that regard, \citep{Freudling_2024}'s work is closer to our area of research. In this paper the authors propose a data processing system that assembles processing tasks into workflows for different use cases \textit{based on pre-existing flowcharts defined by scientists}. An end user only needs to select the use-case and the EDPS system navigates the flowchart to assemble the series of tasks needed to process the data into a workflow. In contrast, our research is focused on discovering the decision tree from which flowcharts can be created. Conceptually, the output of our system (example Fig. \ref{fig:dtc-2step}) could lead to a specification of conditional processing steps (flowchart) which a system like EDPS could follow to assemble a workflow in real time. An interesting connection for potential future research would be to model our output in a form that may be more directly usable by such systems.

Other instances of Reinforcement Learning and Q-learning applied to astronomy include telescope scheduling for observations \citep{terranova2023}.

\section{Simulated Data} \label{sec:sims}
For our study we use simulated data so we may control an emitting source's spatial structure as well as the properties of the corrupting gain signal. This allows us to constrain the complexity (number of actions and input features) of the problem to a simpler regime which requires a very limited set of calibration and imaging options. For simulated data, the hierarchy of organization is as follows: the fundamental data component is the visibility amplitude which is the amplitude of the interference function representing a given spatial frequency \citep{2017isra.book.....T}. In a typical observation, many thousands of visibility amplitudes are measured from the many baselines between antennas as well as across frequency channels and time. In practice this conglomerate of visibility amplitudes is called as a Measurement Set but for our purposes we will refer to the generic name of dataset as we have simulated these quantities. We have created thousands of these datasets which makes up a population and we have created two populations of datasets: one population containing RFI contamination and one without. The reasons for making these populations will become clear in the later sections.

At the individual dataset level, we simulated the visibility data by mimicking the VLA D-configuration array of 27 antennas, forming 351 unique baselines. This configuration `observed' an artificial sky for 10 hours. We limited the observing setup to have 100 frequency channels and 100 time integrations. The artificial sky that was being `observed' consisted of a 1Jy point source object at the phase center to replicate looking at a calibrator source with a flat spectrum. From this starting set of visibilities we introduced sinusoidal signals to represent direction-independent antenna gain distortions that would need to be corrected for by calibration.

The first population of simulations consists of datasets with gain distortions that vary in mean value, amplitude, and period for both frequency and time dimensions. Antenna-to-antenna offsets were also applied to be more realistic. The noise level was kept constant across the population. For the second population we included RFI contamination in addition to gain variations. We chose to include a specific type of RFI, broad antenna-based RFI, because this type of RFI contamination will distort the calibration corrections if it’s not removed before calibration procedure takes place. To be considered "broad", the RFI contamination covers multiple time integrations and frequency channels with very high amplitudes. For each simulation in this population, the number of RFI contaminants, the amplitude of each contaminant, and location in the time and frequency varied.

\begin{figure*}[hbt!]
    \centering
    \includegraphics[width=\linewidth]{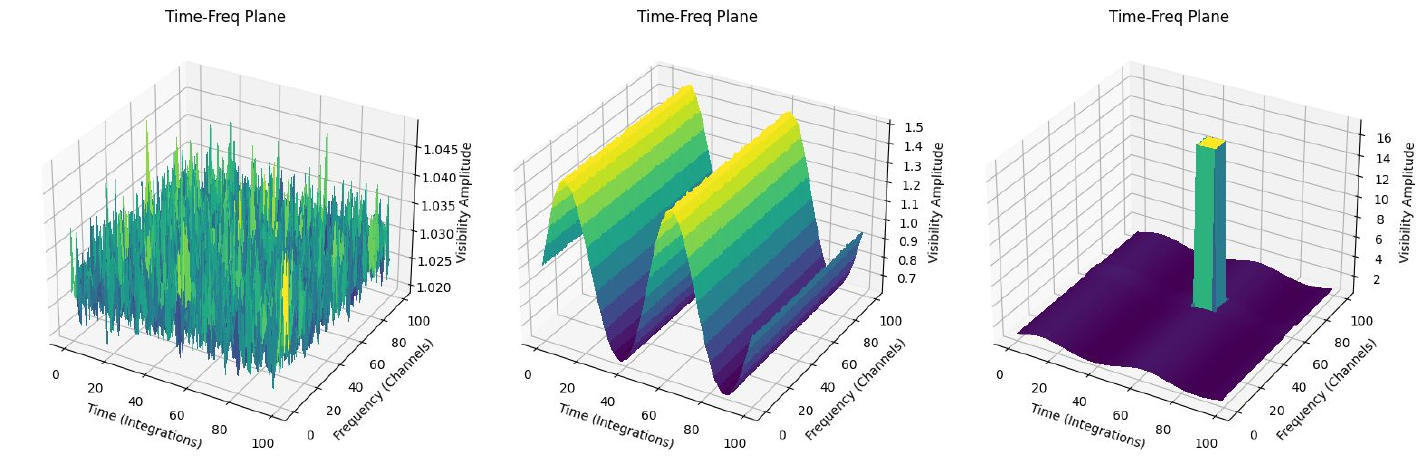}
    \caption{Visibility amplitudes plotted in 3D for one baseline. Other baselines in the dataset are similar in shape but vary in intensity according to the antenna-to-antenna variations. X-Y axes form the time-frequency plane with visibility amplitude on the z-axis (also indicated by color). Left: A simulation without any gain distortions, which is a flat time-frequency plane with noise. Center: Visibilities containing a large gain distortion that varies along the time axis that would require calibration corrections for each timestep. This simulation would benefit from averaging along frequency prior to calibration. Right:  A simulation dominated by an RFI outlier that would need to be removed before estimating any potential underlying gain distortions.}
    \label{fig:tf-planes}
\end{figure*}

{Figure \ref{fig:tf-planes} shows some example dataset visibility amplitudes. For each plot the z-axis is visibility amplitude and the x and y axes correspond to the frequency and time dimensions. A perfect instrument, ignoring noise and atmospheric effects, observing a flat-spectrum point source at the phase center would be a constant plane in time and frequency at the mean visibility amplitude. In our case this would be 1 for the 1Jy source simulated (Fig.\ref{fig:tf-planes} Left). Any distortions from antenna gains or outliers will result in some distorted shape in the plane (Fig.\ref{fig:tf-planes} Center and Right).

\section{The Decision Environment} \label{sec:p2}
To demonstrate this application we focus on a specific decision in the calibration stage, namely, the choice for the amount of averaging to apply (if any) before removal of the antenna gains introduced by the instrument. The purpose of calibration is to estimate the gain distortions and remove them, leaving the time-frequency plane effectively flat, meaning the visibilities measurements are unaffected by the instrument. 

To make a data-driven decision we first need to define the context for making a decision. We start by deciding what properties of the dataset are needed to adequately describe the state of the data, what actions are available to choose from, and by defining the objective. The objective is used to form a metric to steer decisions.

\newpage
\subsection{Data State: The Input Features} \label{sec:p2.1}
For describing the state of the dataset we use four statistical properties measured across all baselines, referred to as the state-vector. 

\begin{enumerate}
    \item The mean visibility amplitude, where a value other than the known flux of the calibrator point source is an indicator of a need for calibration. In this paper, we also refer to this quantity as 'mean gain' because our simulations use a 1 Jy point source.
    \item Standard deviation of visibility values along the frequency axis
    \item Standard deviation of visibility values along the time axis 
    \item Maximum visibility amplitude
\end{enumerate}

We identified the four quantities listed above as the minimum required for the controlled environment of our experiment. Our sky structure is a single point source at the phase center with all baselines measuring the same value, and all deviations from a flat visibility function are due only to antenna gain effects. For our limited set of actions, we only need to distinguish between time versus frequency variability in the data, mean visibility amplitude, and the existence of outliers that may require flagging. The standard deviation values indicate if there are preferential axes along which SNR may be improved by averaging. The presence of outliers would also be captured in the mean value, standard deviation(s), and max value. For example, the state-vector for Fig. \ref{fig:tf-planes} Left is [1, 0, 0, 1]. For Fig. \ref{fig:tf-planes} Center it is [1, 0, 1.1, 1.4], and for Fig. \ref{fig:tf-planes} Right it is [1.4, 0, 1.1, 12].

This simplification allows us to carefully validate our approach and unambiguously interpret our results.  In subsequent work, when we include source structure, we will need more refined metrics to distinguish genuine baseline-to-baseline variations from antenna gain effects. In a real-data environment, where there may be more effects to account for, the dimensionality of the state vector will undoubtedly increase, possibly to include image statistics as well.

\subsection{Choices: Actions Available} \label{sec:p2.2}
The actions represent the possible choices one can make in a decision. We selected actions that are relevant for the decision scenario at hand, which are typical to what an end-user would choose from. The actions we made available to choose from are the following:

\begin{enumerate}
    \item Action 1 - average the frequency axis and calibrate. This is best suited for when a dataset has minimal variation across frequency but has variation in the time axis. Averaging the frequency axis increases SNR for the more variable time axis which can improve the calibration corrections. The calibration runtime is reduced by averaging one of the axes. 
    \item Action 2 - average time axis and calibrate. This is best suited for when a dataset has minimal variation across time but has variation in the frequency axis. Averaging the time axis increases SNR for the more variable frequency axis which improves the calibration corrections. The calibration runtime is reduced by averaging one of the axes.
    \item Action 3 - average both the time and frequency axes and calibrate. This is best suited for when a dataset has minimal variation in both time and frequency and averaging both axes would have negligible impact on calibration quality.  The calibration runtime is further reduced from averaging both axes.
    \item Action 4 - average none and calibrate. This is best suited when a dataset has variation in both time and frequency. No averaging is applied as any averaging would result in poor approximations and therefore poor calibration corrections. This is the most compute intensive as the data size is not reduced from averaging.
    \item Action 5 - do nothing. This is for the case that your dataset has no outliers to remove or gains to correct (it may already be calibrated). No calibration corrections are calculated or applied. This takes zero runtime.
    \item Action 6 - flag RFI (outliers). This is best suited when RFI is present in the dataset. The flagging algorithm removes outlier data based on sliding window measurements. This takes a static runtime.
\end{enumerate}

The actions described here apply to all baselines in the given dataset, not a single baseline. The actions described here are implemented using the CASA software \citep{The_CASA_Team_2022}. The CASA software allows for one to calculate the corrections without having to apply them unless explicitly directed with another function. For convenience, we have combined the creation and application of corrections here with Actions 1-4.

\subsection{Objective and Subsequent Metric} \label{sec:p2.3}
Our objective is to remove any effect on the visibility data from the antenna gains and outliers. How we measure any unwanted contaminants is by looking at the distribution of residuals after model reconstruction of the image. The ideal residuals in the image domain are a noise-like Gaussian distribution at the theoretical noise level (set during simulation). This forms part of the metric we use to measure how “good” an action is. Our metric is based on two factors:  

\begin{enumerate}
    \item After an action is taken, how close is the residual distribution to the theoretically predicted noise distribution. Incomplete or improperly estimated calibration corrections and unflagged outliers can change raise the residuals above what is theoretically possible. 
    \item The runtime it took for the action to achieve those residuals. This allows us to consider the runtime property associated with actions; something a human intuitively does but is not quantified in current pipeline heuristics.  
\end{enumerate}

\subsubsection{Image Quality by Comparing Residual Distributions}
The distribution of image-domain residuals is used to evaluate the quality of actions that were applied. To compare the two distributions, a Savitzky-Golay \citep{doi:10.1021/ac60214a047} filter is applied to a finely binned histogram of the pixel amplitudes in the residual image as well as the known theoretical distribution. This filtering results in a smoothed histogram that is less sensitive to noise and represents the distribution shape. The smoothed values at each bin are compared between the two distributions and a Wasserstein metric, or Earth Mover's Distance (EMD) \citep{rubner2000earth}, is calculated (Fig \ref{fig:dist-comp2}).

\begin{figure}[hbt!]
    \centering
    \includegraphics[width=1\linewidth]{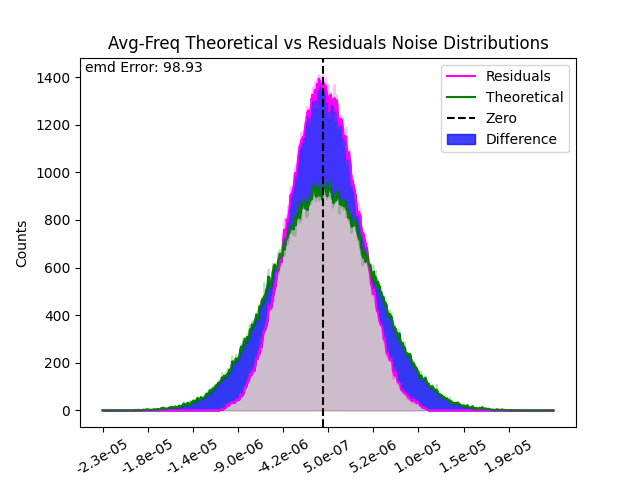}
    \caption{Figure: An example comparison between theoretical noise distribution (green) and a model’s residuals distribution (red) after averaging both axes before calibration. The faded magenta and green distributions are the original sampled values while the solid magenta and green lines represent the distributions after the Savitzky-Golay filter is applied. The blue highlights the difference between the smoothed distributions from which the EMD is calculated, indicating correctness (lower is better). The vertical dashed line indicates where the 0 point is; theoretical noise is centered around 0.}
    \label{fig:dist-comp2}
\end{figure}

This metric is sensitive to distribution shape and its distance from ideal. It responds well to the presence of imaging artifacts and other error patterns as well as differences in Gaussian random noise levels. It is also agnostic of the source structure. For these reasons, we chose this metric over typically-used metrics of standard deviation of the residuals or image fidelity. Figure \ref{fig:actions-avg-freq} shows an example of a simulation with each of the actions applied and their respective results. 

\newpage
\subsubsection{Runtime}
The runtime of the calibration step is estimated as being proportional to the number of elements presented to the calibration solver, where an element is a distinct time-frequency portion. The calibration solver performs a non-linear least squares calculation on each element. The time it takes to compute corrections changes based on the amount of averaging applied to the input data as that changes the number of elements presented to the calibration solver. Note that while averaging may be used on the input data for the solver, it does not change the original time-frequency plane. If the data are averaged, the shape is changed for the input data to the calibration solver but the resulting corrections are applied to each element in the original time-frequency plane. For example, the 100 channels and 100 integrations is a 10,000 element time-frequency plane (100x100). If we choose to average one axis, the data shape changes for what is given to the solver (1x100) but the corrections are applied to the original 10,000 element plane (100x100). We use a runtime approximation based on the time it takes to complete a series of additions versus completing an iterative non-linear solver for an arbitrary size. From that we assume non-linear solving is approximately 80x slower than averaging data. The runtime is calculated as: \[ (N_{elements-averaged} * Cost_{avg-sol}) + (N_{elements-solved} * Cost_{sol}) \]

Therefore, based on the size of our datasets and the approximation used, the runtime cost of each action is as follows:
\begin{itemize}
    \item Action 1, average frequency axis, has a cost of (9900 * 1) + (100 * 80) = 17,900
    \item Action 2, average time axis, has a cost of (9900 * 1) + (100 * 80) = 17,900
    \item Action 3, average both axes, has a cost of (10000 * 1) + (0 * 80) = 10,000
    \item Action 4, average none, has a cost of (0 * 1) + (10000 * 80) = 800,000
    \item Action 5, do nothing has a cost of 0
    \item Action 6, flag outliers, has a flat cost of 1,000,000
\end{itemize}

These values are intentionally hypothetical to exaggerate the runtime differences between actions to test how our proposed solution would take these values into account.

\subsubsection{Combining Image Quality and Runtime into a Single Metric}
The EMD and runtime values for an action are summed to give a single numerical measure of performance to represent how good (or bad) that action was to take on the given dataset. This summed value is the action-value. We can evaluate each individual action this way on a given dataset. The longer the runtime is the higher the runtime penalty and the higher the EMD value is the higher the image quality penalty, which combine to form a high action-value. Therefore in our scenario, the action with the lowest action-value is considered the optimal action. 

With the EMD value typically being much smaller than the runtime value, the raw runtime value dominates the result and therefore we need a way to equalize the scales of the two values. In our implementation, a scalar of 1e6 is applied to the raw EMD value for comparative scaling. In general, this scalar provides a convenient way to give preference to one factor over the other, such as prioritizing runtime over image quality in the action-value statistic. 

\section{Data-Driven Action Selection}
With the datasets simulated and the decision environment defined (state-vectors, actions, metrics, and objective function), we can now evaluate how different actions perform on a given dataset (state). In the case of a single decision we select the optimal action based on how well the action performs towards the desired objective, as measured by our metric. Figure \ref{fig:actions-avg-freq} shows an example dataset that contains gain distortions along the time axis of the left hand side. This is an example where there is sufficient signal-to-noise to perform accurate calibration on each time and frequency sample, but we expect that averaging along frequency before calibration will provide similarly good numerical accuracy at a much lower runtime cost.

\begin{figure*}[ht!]
    \centering
    \includegraphics[width=1\linewidth]{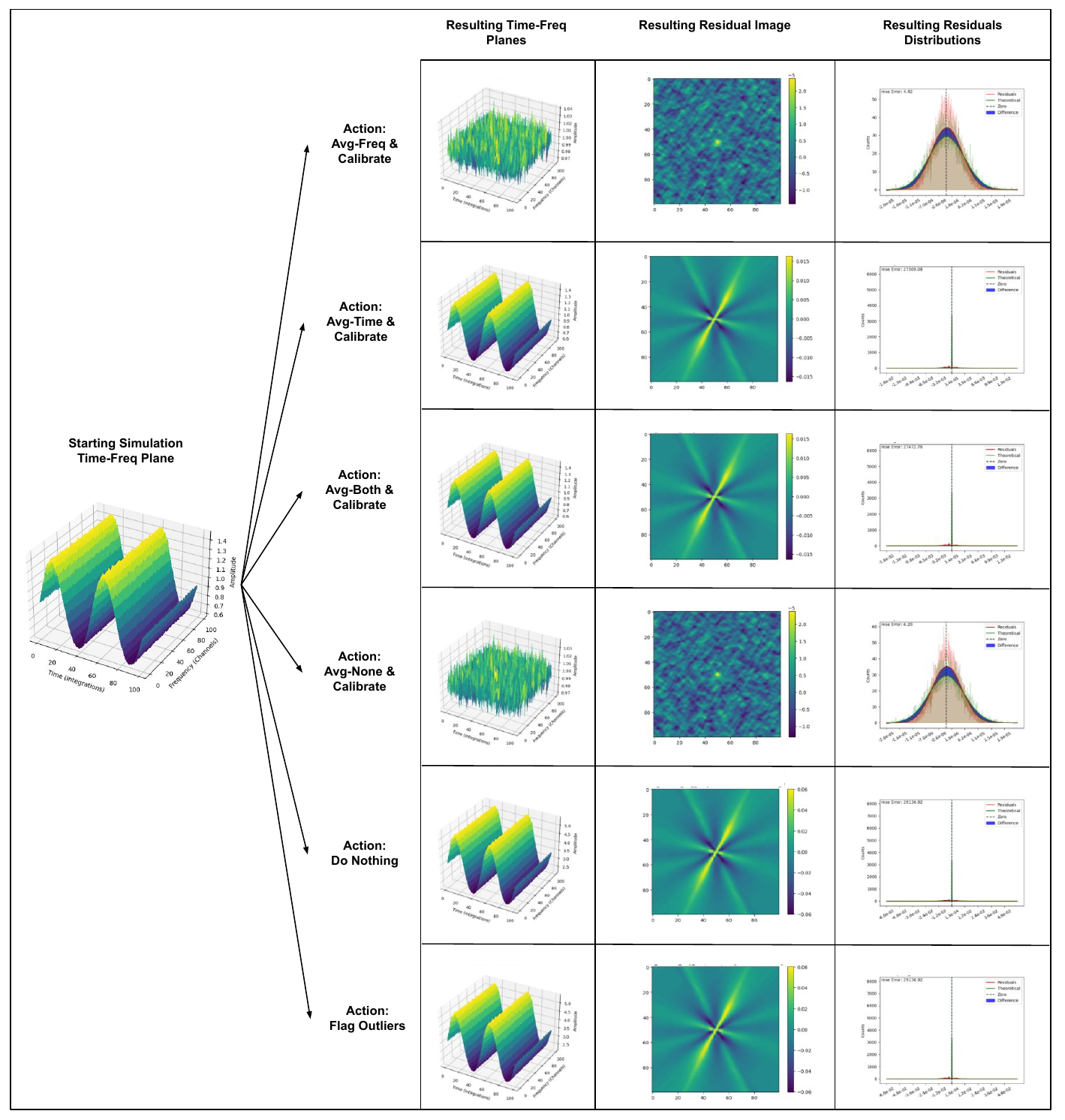}
    \caption{Leftmost: Plotted visibilities of a simulated dataset containing gain distortions.  Branching from that simulation is each possible action and the results each action has on the data. Each row of images associated with each action (from L→ R) are the resulting time-frequency plane, the image residuals, and the residuals distribution compared to the theoretical noise distribution. Color scales and X-Y axes values are synchronized on all plots to make relative visual comparison easy. Image residuals that look noise-like, and residual distributions that look Gaussian, indicate better numerical outcomes such as row 1 and 4. Rows 1 and 4 have RMS levels of 1e-5Jy/beam whereas rows 2, 3, 5, and 6 have an average of 3e-2Jy/beam.}
    \label{fig:actions-avg-freq}
\end{figure*}

Each action is independently applied to this dataset and evaluated. From this diagram it is easy to see there are two actions, row 1 and 4, that give us the desired result of removing the gain distortion from the time-frequency plane. The "incorrect" actions in this scenario, rows 2, 3, 5, and 6, result in much worse residual RMS levels at an average of 3e-2 Jy/beam with structure still present in the residual image. Row 1 and 4 correspond to Actions 1 and 4. Action 1 is averaging the frequency axis before calibrating and Action 4 does no averaging before calibration. The residuals values between these two actions are similar with RMS levels at 1e-5 Jy/beam and would result in a tie if only considering imaging performance. However, Action 4 takes 44x longer to finish than Action 1. This demonstrates why we consider runtime in our action-value metric. With our action-value metric taking into account runtime and image quality, Action 1 is the clear choice. 

While Fig \ref{fig:actions-avg-freq} is one example, with a small number of actions and a finite number of simulated datasets, it is feasible to maintain a table that contains the action value for each state-action pair. This table forms the basis for tabular methods in Reinforcement Learning. Tabular methods are simple yet foundational to understanding the more advanced Reinforcement Learning methods. We use the table of known values, a Q-table, as a basis for Q-learning, a Reinforcement Learning method that learns the optimal action to take in a given state \citep{watkins1989learning} \citep{watkins1992q}. 

Q-learning is necessary for learning non-greedy behavior, especially if an action’s effects only manifest later in the decision chain. We intentionally created a population of simulations contaminated with RFI to test solving for the optimal sequence of actions. RFI introduces a sequencing problem of not only if but when to flag data in relation to other actions. We want to test whether the correct \textit{sequence} of actions can be found to get to the global minimum. 


\section{Results} \label{sec:p4}
This section will cover the validation of our decision environment and how well Q-learning was able to find the optimal sequence of actions. Population 1 datasets were used to evaluate how well our selected input features and action value metric were at capturing what a domain expert would do faced with same decision. This validation allowed us to use the same environment setup in more complicated scenarios for Q-learning. Population 2 datasets required solving a sequence of actions; we will review the performance of the Q-learning at finding the optimal sequence of actions based on data characteristics themselves (no hardcoded recipes).

To help convey ideal actions across an entire population of datasets we construct a visual aid in the form of a 3D scatter plot, called the state-space. The axes for the 3D space are a sub-selection of the input features of the simulations. The state space allows us to plot entire populations of datasets and view their different properties along with the ideal actions to take on them and the subsequent action regions that are formed. Figure \ref{fig:full-state-space} is an example of a 3D state space with different frequency variations, time variations, and mean gains. By colorizing the plotted simulations with their best-performing action we can see which actions occupy different regions of state space.

\subsection{Validating the Environment with Population 1}\label{sec:p4.2}
Population 1 was designed that each dataset only needed 1 action applied to bring it into an ideal state. The population contains several thousand datasets, each differing in frequency variation, time variation, mean gain, and consequently max value. For each dataset every action was independently applied to determine its action value. Figure \ref{fig:full-state-space} illustrates the simulations in the state-space colorized by their respective best performing action.

\begin{figure}[hbt!]
    \centering
    \includegraphics[width=0.5\linewidth]{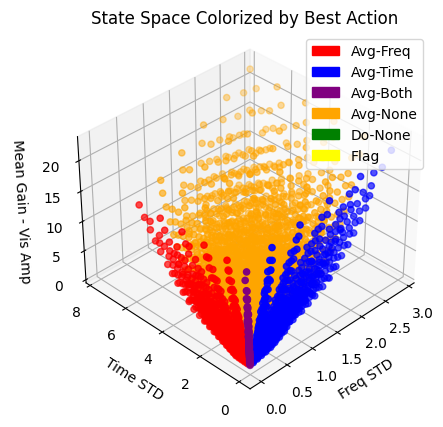}
    \caption{3D state-space with thousands of simulations colorized by best the performing action. State space axes are frequency variation, time variation, and mean gain. The individual points are simulations with those statistical properties. The points are colorized by the best performing action, as determined by our metric. We can see regions in the state space that correspond to ideal actions. Flag action will come later.}
    \label{fig:full-state-space}
\end{figure}

Figure \ref{fig:full-state-space}  shows there are clear and distinct regions within the space. For example, consider all the blue points: these simulations have a low gain distortion along the time axis but a high enough mean gain to require calibration. The optimal choice there is to calibrate after averaging in time. Conversely, the red points have low gain distortion along the frequency axis but a high enough mean to require calibration. The optimal choice there is to calibrate after averaging in frequency. The purple points indicate these simulations can have both axes averaged before calibration without degrading calibration corrections. On the other hand, orange points indicate simulations that should not have any averaging applied before calibration. Any green dots represent simulations that don’t require calibration (not shown) and would reside at the 0 point of all 3 axes. This would represent when there is little-to-no variation and the mean gain is close to the expected value. 

Datasets with similar statistical properties are found to have the same optimal action needed to rectify them, creating the colored regions in Figure \ref{fig:full-state-space}. This is the most basic validation of the input features, metrics, and objective function in our simplified environment. \textit{The regions associated with a particular action match what a domain expert would do in the same scenario, validating our environment and metric.} This gave us confidence to build more complex decision scenarios requiring multiple actions to be tested with the same environment setup.

\subsection{Q-Learning a Sequence of Actions with Population 2}\label{sec:p4.3}
We created Population 2 datasets by adding strong outliers to the existing datasets in Population 1. These outliers simulate broadband, antenna-based RFI corruptions in an observation (Fig \ref{fig:tf-planes} RHS). With these outliers, we now have simulations that warrant the flagging action. However, because these corruptions are placed on top of the existing gain patterns, both the RFI and gain patterns need to be corrected for. Importantly, this type of RFI will distort the calibration solutions if they are not removed \textit{before} calibration takes place. This introduces a new problem to solve which is finding the optimal action sequence.

For a simulation that has gain distortions and RFI at the levels we have chosen to add, a greedy algorithm will opt to do gain corrections first since that has an immediate stronger effect on image residuals than flagging RFI. This is because the gain distortions affect more of the time-frequency elements than sparse RFI, contributing more weight to the statistical properties of the dataset. However, RFI distorts the gain corrections resulting in a sub-optimal calibration. This makes the sequence of actions critical with Flagging needing to be run before gain corrections (if applicable). A greedy algorithm will not select the ideal sequence of actions and will not reach the global minimum. As described earlier, Q-learning is a way to learn a non-greedy approach for sequencing actions.

Q-learning iteratively updates the action value in a given state-action pair as new combinations of states and actions are explored. In our scenario for Population 2 there are two decisions (steps) to be made with six actions possible for each step, totalling 36 possible permutations (with replacement) of actions. For example, step 1 could be Flag and step 2 could be Avg-Freq and calibrate. As different pathways are explored the action values for any single action is updated based on final outcome, known as the total reward. The total reward is a value composed of the \textit{cumulative} runtime of all actions in the pathway and the \textit{final} image quality; intermediate image results are not considered. The total reward is then used to update the individual action values of the actions making up the pathway. This way the actions are affected by the final results they create, good or bad.

An example of the evolution of action value estimates for the first action to take on a dataset with RFI can be seen in Fig. \ref{fig:value_evolution}. As the Q-learning algorithm explores different action sequences the final reward value influences the preceding action's value. With enough iterations to sample different sequences of actions, the values for each action converge, revealing the optimal action to take in the first decision (flag).}

\begin{figure}[hbt!]
    \centering
    \includegraphics[width=0.75\linewidth]{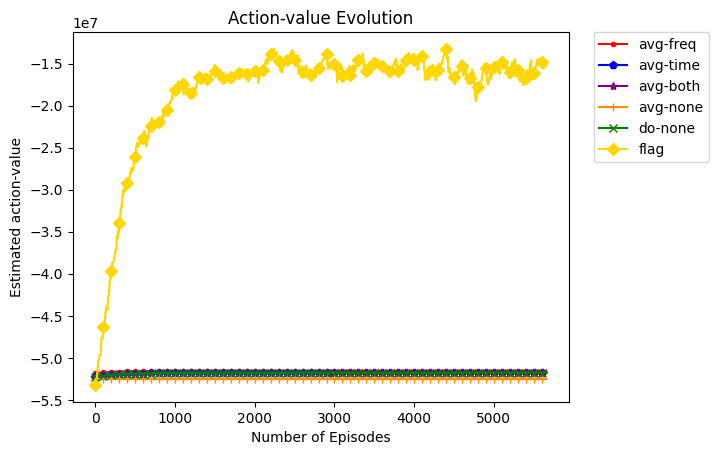}
    \caption{Action value evolution for a dataset with RFI contamination. After a few iterations of updates to the action-values the `Flag' action rapidly becomes the most rewarding action to take.}
    \label{fig:value_evolution}
\end{figure}

\begin{figure}[hbt!]
    \centering
    \includegraphics[width=0.75\linewidth]{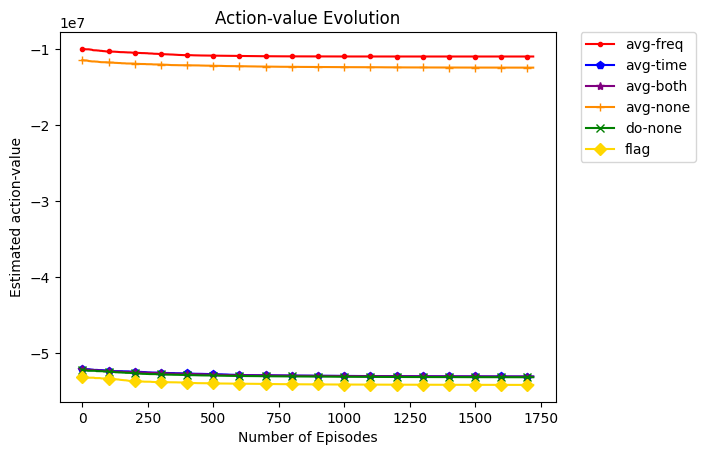}
    \caption{Action value evolution for a dataset without RFI contamination. `Flag' is not the ideal action to take but Avg-Freq and Calibrate.}
    \label{fig:value_evolution2}
\end{figure}

For comparison, the Q-learning method was applied to the same dataset once the RFI was removed. Figure \ref{fig:value_evolution2} gives an example of the action-value evolution in this state. Q-learning found Flagging was no longer a rewarding action to take but to Avg-Freq and calibrate directly.

\begin{figure}[hbt!]
    \centering
    \includegraphics[width=0.5\linewidth]{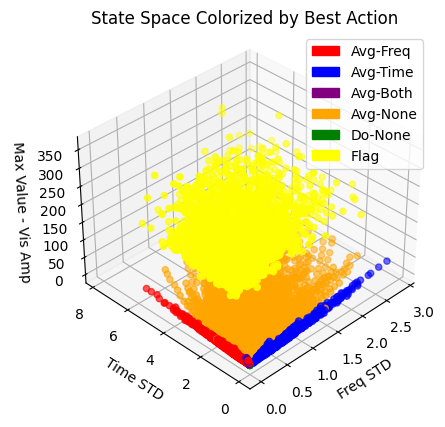}
    \caption{3D state-space of Population 2 colorized by best the performing action.}
    \label{fig:rfi_state_space}
\end{figure}

\newpage
We applied the Q-learning algorithm to each dataset in Population 2. Plotting each of these simulations in a 3D state-space, colorized by the optimal action to take gives Figure \ref{fig:rfi_state_space}. 

Note that since these datasets required two actions, when the first action was applied to the dataset it changed its statistical properties resulting in a new state of the data. The resulting state of the data after the first action is also plotted in the state-space resulting in more points. These are intentionally plotted so the original regions from Population 1 can be recognized. The yellow "point cloud" is comprised of the same datasets as Population 1 but contaminated with RFI. Q-learning has estimated the Flag action to be the optimal action to take first; if this was a greedy approach the actions and therefore colors would be different.


It is important to highlight that although we have only applied Q-learning to two decisions here, this method can be applied to a sequence of any length.

While the Population 2 datasets provide discrete data points across a wide range of values in the state-space, it would be beneficial to learn a continuous function that provides an optimal action at \textit{any} location in the state-space so that we can create a direct map from an input dataset's characteristics (via its state vector) and the optimal action. To that end, we chose two different methods to model the action regions in the state-space, a simple neural network and a decision tree classifier (DTC). 

\subsubsection{Modeling the Action-Regions with a NN}\label{sec:p4.3.4}
Neural Networks (NN) are able to model any continuous function to an arbitrary precision (universal approximation theorem, \citep{hornik1989multilayer}) and have become a standard tool for such applications. In short, the fundamental building blocks of a feedforward NN can be thought of as a collection of piecewise elements that can be (hierarchically) combined to replicate more complex functions. The contribution of any individual building block to the overall function is determined by a weight which is optimized through standard optimization techniques.  We chose to test NN for their ability to handle high dimensional inputs and modeling non-linear functions, which will become typical when applied to more complex decision environments and real data. 

\begin{figure}[hbt!]
    \centering
    \includegraphics[width=0.5\linewidth]{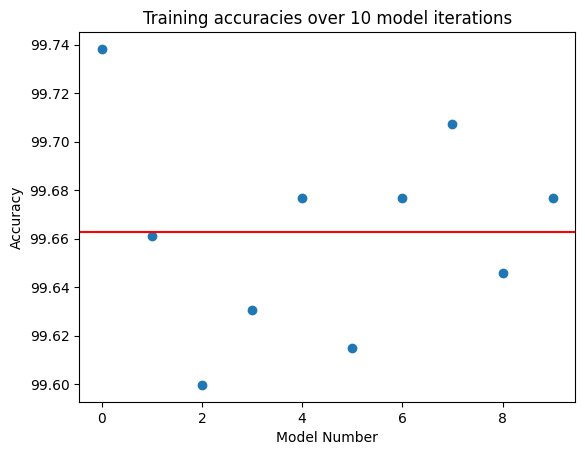}
    \caption{10 independently instantiated NN models with prediction accuracy of each model as a blue dot. The red line is the 10 model average. On average, a trained model's prediction accuracy on unseen datasets is $>$99.6\% accurate.}
    \label{fig:nn-2step}
\end{figure}

\begin{figure}
    \centering
    \includegraphics[width=0.5\linewidth]{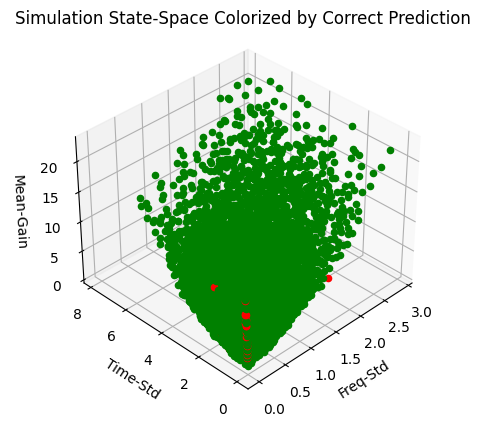}
    \caption{Trained NN model at predicting the ideal action for each dataset in the population. Correct predictions in green, incorrect predictions in red.}
    \label{fig:nn-predictions}
\end{figure}

The input data was the Population 2 datasets with RFI. The input features were the same four statistical measurements (mean gain, frequency variation, time variation, max-value) and the ideal action was the output. We trained the NN on a subset of the data and used the remaining unseen datasets to validate the NN model. Since there is inherent variability in NN modeling from the weight initialization and (mini-batch) stochastic gradient descent, one is unlikely able to recreate NN models with the same prediction accuracy. To that end, we trained 10 models from scratch and took the average value, measuring ~99.6\% in prediction accuracy, as shown in Figure \ref{fig:nn-2step}. An example of one of the models action predictions given the state vectors for all the simulations in Population 2 shown in Figure \ref{fig:nn-predictions}.

NNs are powerful at modeling continuous functions especially as complexity grows. However, interpretability of what the NN is learning becomes more difficult. We reserve applying and developing interpretable methods for future work.

\newpage
\subsubsection{Modeling the Action-Regions with a DTC}\label{sec:p4.3.3}
A decision tree classifier (DTC) is a simple, easy to understand way to find regions of the state space associated with different ideal actions. Crucially, the DTC can give the explicit rules behind its regions which gives interpretability to the method and makes it simple for checking alignment with domain experts.

\begin{figure*}[hbt!]
    \centering
    \includegraphics[width=0.75\linewidth]{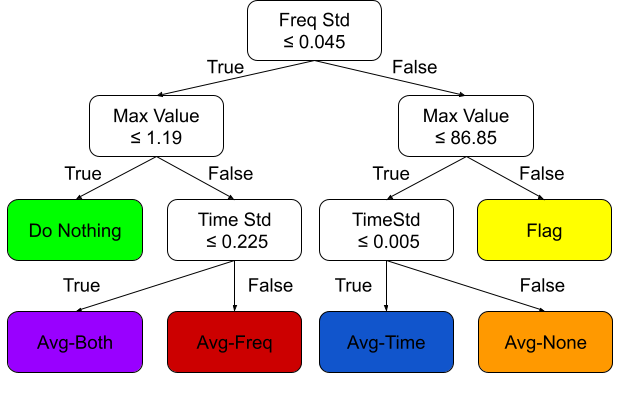}
    \caption{A decision tree and its rules based on the state-space. The DTC model gives us interpretability to the boundaries of the different action-regions and shows alignment with domain expertise. In each of the leaf nodes of the tree we see the unique actions and the DTC’s ability to find numerical boundaries that best partitions the data. }
    \label{fig:dtc-2step}
\end{figure*}

We fit a DTC to the ideal actions for Population 2 datasets to extract the numerical rules for the different action regions. While the 3D state-spaces only visualize 3 axes of the data at once (for clarity), the DTC has access to all the input features of the data for deciding where to split the data. Figure \ref{fig:dtc-2step} shows the model found by the DTC.

The DTC model gives us interpretability to the action-regions and shows alignment with domain expertise. In each of the leaf nodes of the tree we see the unique actions and the DTC’s ability to find numerical boundaries that partition the data. A learned DTC can be used to process an unseen dataset simply by using the statistical properties of the dataset (state-vector) and following the decision tree's workflow for each step.

\subsection{Adaptation to Software Changes}\label{sec:p4.4.3}
Software evolution is the continual development of software after its initial release to address problems or changing requirements. Astronomy software, and CASA in particular, is not immune to this. Different software tools and algorithms are necessary for handling different use-cases, observing modes, hardware changes, telescope designs and so on. In some cases, different software tools can be used to accomplish the same goal but with differing accuracies and efficiencies. If a new software tool is created, or a fix is made to an existing one, there is a practical need to try and evaluate drop-in replacements. The RL approach outlined here, estimating action-values and finding optimal pathways, works regardless of whether the tools change. It finds the optimal processing path with the tools it is given. 

\begin{figure*}[hbt!]
    \centering
    \includegraphics[width=1\linewidth]{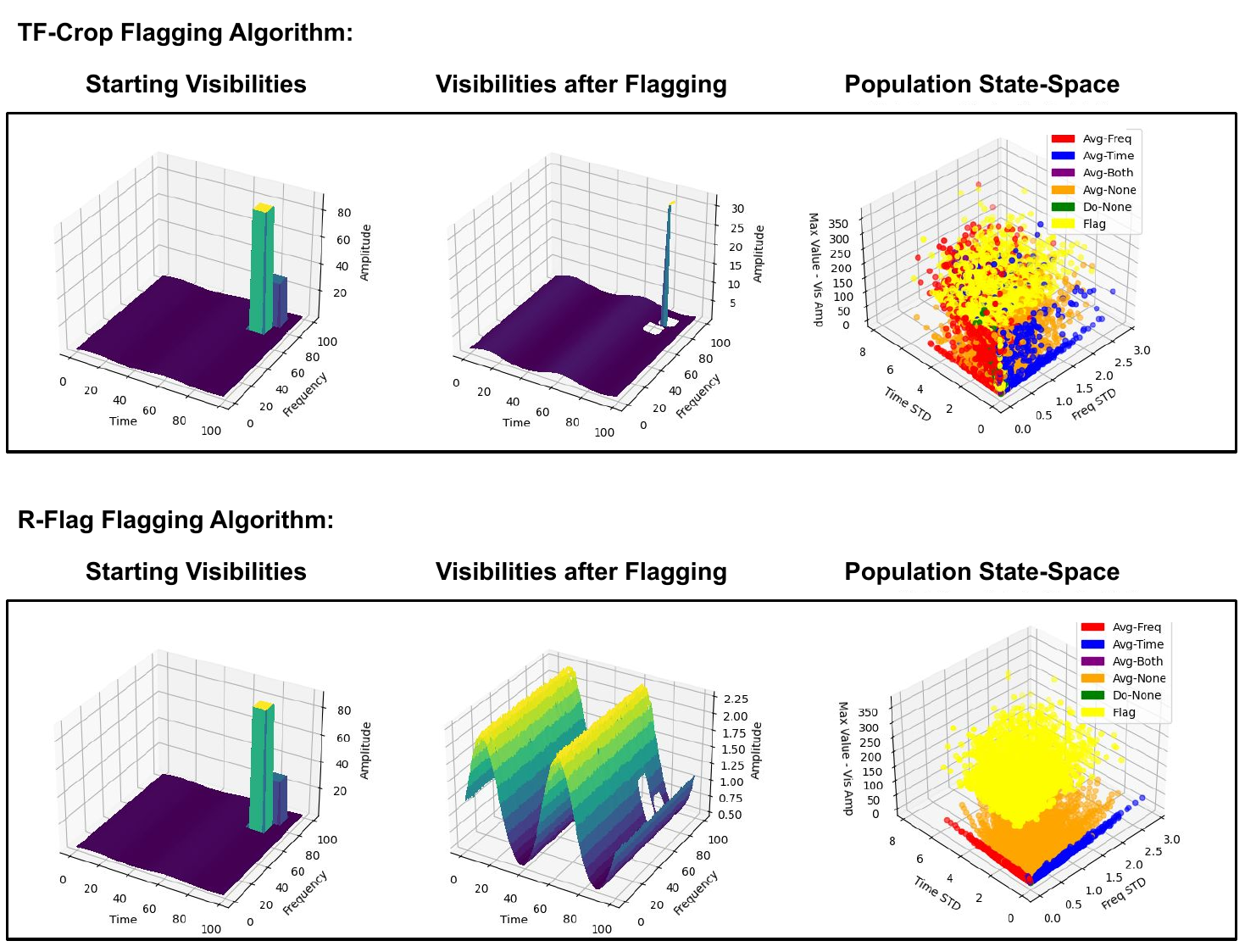}
    \caption{Comparing performance of two different automatic flagging algorithms available in CASA. The top panel is the TF-Crop algorithm and the bottom panel is the R-Flag algorithm. Both algorithms start with the same initial dataset (L) and the resulting Time-Frequency plane after flagging is applied is shown in the middle column. The right column shows the resulting 3D state-space of all datasets contaminated with RFI when processed with the given algorithm. Distinct regions in the state space indicate better performance at the given task. }
    \label{fig:tools}
\end{figure*}

This is demonstrated by using the two different automatic flagging algorithms that exist in CASA. These automatic flaggers have their respective use cases and were designed for as such, meaning their accuracies and efficiencies differ. We used them with their default parameter settings to highlight the differences between them. The RFI we added to the datasets is easily identified and removed by one algorithm (R-Flag) and not the other (TF-Crop).  We tested all the contaminated datasets with both of the R-Flag and TF-Crop flagging algorithms as the Flag action. Since the performance of each flagging algorithm is different, the optimal pathways found for each dataset are different. This should be of no surprise, if the tool’s performance is different than the resulting data states are different, potentially leading to different actions downstream (Fig. \ref{fig:tools}). 

The top panel is the TF-Crop algorithm and the bottom panel is the R-Flag algorithm. Both algorithms start with the same initial dataset (left column) and the resulting time-frequency plane after flagging is applied is shown in the middle column. You can see for the TF-Crop algorithm, set with default parameters that are known to not be robust to broadband RFI, there are outliers still remaining after flagging. These leftovers continue to contaminate the data. For R-Flag, the algorithm is able to correctly identify and remove all outliers, therefore exposing the gain distortions that are present. The rightmost column shows the 3D state space of all the contaminated simulations colorized by their ideal action. For TF-Crop the leftover outliers caused data states to exist that could not be rectified by flagging alone. As flagging would not improve the dataset any further, the best action was to calibrate without any averaging to get the dataset as-close-to-ideal-as-possible given the tools available. An example is shown in the next section. When the population of datasets with RFI was processed with R-Flag the action-regions in the state space were much more apparent and organized. This indicates R-Flag working accurately to remove \textit{this type} of RFI on the first attempt.

Learning when, or when not, to use a certain tool in a situation is the subject of training workshops for end-users and the work of scientific developers in observatories. RL automates the heuristic discovery of what tool(s) to use and when, removing human overhead. Secondly, the action-values themselves can be a powerful metric for investigating the performance of the different algorithms and software tools available.

\subsection{Data-Driven Decision to Repeat Flagging}\label{sec:p4.4.2}
As seen in Figure \ref{fig:tools} when R-Flag was applied to Population 2 it was able to detect and completely remove outliers. On the other hand, when using the TF-Crop algorithm, there were some simulations that had RFI contamination leftovers. Those datasets with leftovers provided us a situation to demonstrate Q-learning's adaptable data-driven capabilities and repeating flagging as necessary.

Here we highlight one dataset in particular that has high- and low-amplitude RFI contamination in addition to gain distortions. This is a specific example of TF-Crop flagging the high-amplitude RFI that would affect calibration solutions, but not catching the low-amplitude RFI whose effect would show up only after calibration and still produce image artifacts. We did this to test if Q-learning could learn to repeat an action if and when needed.

\begin{figure*}[hbt!]
    \centering
    \includegraphics[width=1\linewidth]{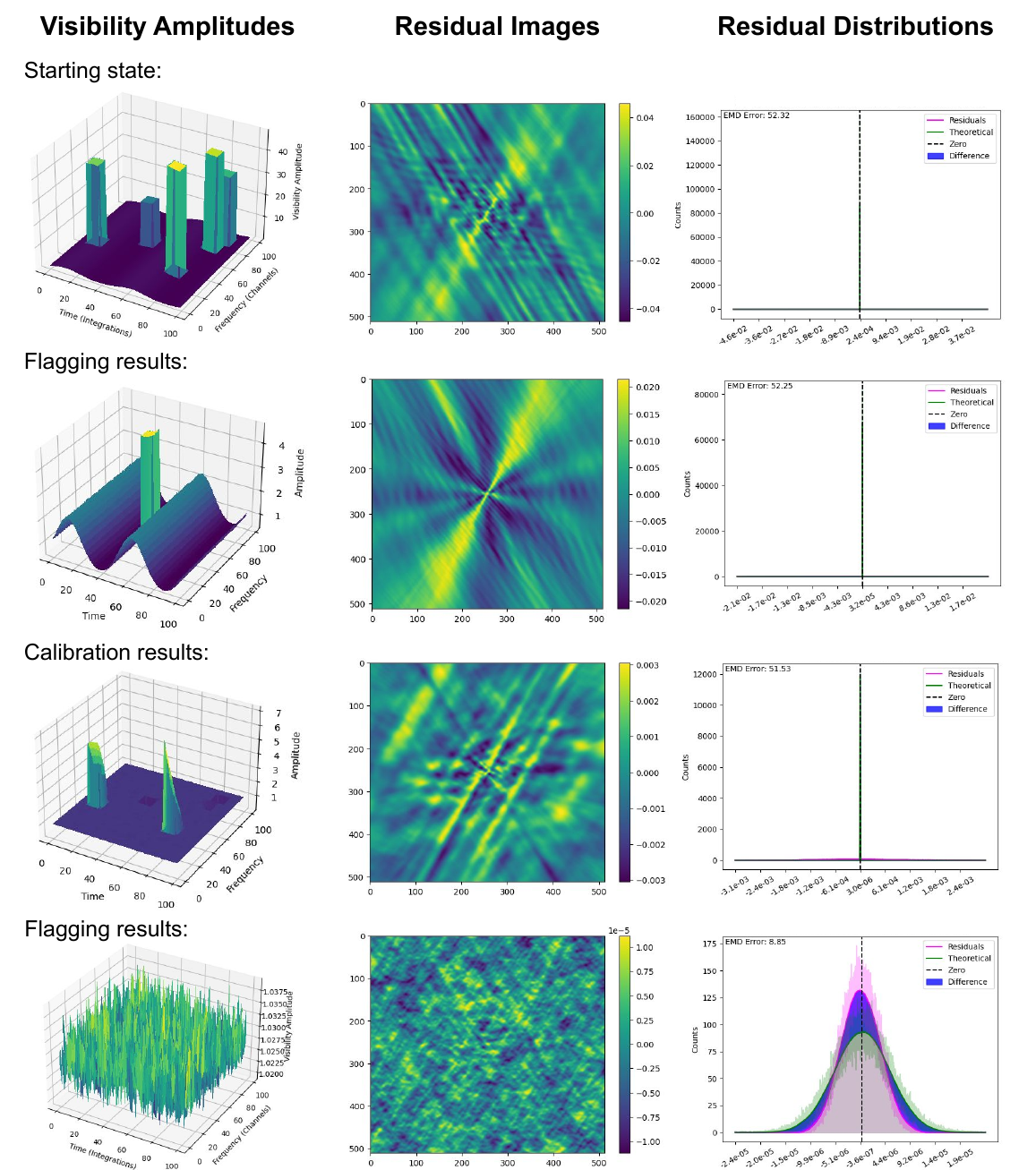}
    \caption{Demonstrating Q-Learning's ability to take actions based on the contents of the data. Columns (L$\to$R) show the visibilities amplitudes in the time-frequency plane, the corresponding (residual) image plane, and the residuals distribution. The progression from top to bottom is the different actions applied to the data. The decision to flag a second time based on the results of calibration is data driven and not from prescribed instruction.}
    \label{fig:flag-cycle}
\end{figure*}

Figure \ref{fig:flag-cycle}'s columns (L$\to$ R) show the visibilities amplitudes in the time-frequency plane, the corresponding (residual) image plane, and the residuals distribution. The top row is a dataset contaminated with RFI. The time-frequency plane shown in this row shows the max values across all baselines to illustrate the amount of RFI present in total. Flagging with the TF-Crop algorithm is then applied and the results are shown in the second row. This time-frequency plane is showing a single baseline only, particularly one that is still contaminated with RFI that was missed by the TF-Crop algorithm at its default settings. Since the amplitude of the remaining RFI was "low enough" the dominant feature from the state vector was now the variations from the antenna gains. Q-learning found the next best action to take was to calibrate (as flagging again would make no difference). The calibration correctly flattened the wavy gain distortion in the time frequency plane but the dataset is still plagued by outliers (the different locations and shapes of outliers are a product of incorrect gain corrections applied to the dataset). Q-learning determined the next best action was now to flag again because the removal of the gain distortions changed the statistical properties of the dataset allowing the low-level RFI to be detected and removed.

This is a contrived example to show the dynamic data-driven decision making ability (in a situation where simply lowering of the TF-Crop threshold was not an option). Without specifying, Q-learning was able to learn staggered flagging actions to remove all the adverse affects. This demonstrates the data-driven capabilities of Q-learning. Flagging multiple times without being explicitly programmed to do so exhibits flexibility beyond what a current pipeline can do. In current operations, this is a situation that requires a human to fix.

\newpage
\section{Discussions}\label{sec:p5}
Currently the standard way to process data is with a predefined, rigid, sequence of processing steps that is designed for the average (or specific) use-case. For the case of ALMA and VLA pipelines, the sequence and steps have been slowly shaped over 10+ years of heuristic development \citep{2023PASP..135g4501H}. In contrast, Reinforcement Learning methods learn functions for mapping situations to actions so as to maximize a numerical reward signal. These methods are driven by the data characteristics themselves, not some predefined recipe. The benefits of this approach would be to achieve large scale automation while simultaneously achieving science-ready quality of the data products. 

\subsection{Data-Driven Processing}\label{sec:p5.1.1}
RL methods provide a true data-driven approach to these problems with flexibility to adapt to the needs of the data. This flexibility provides an advantage that can be used for improving data quality or increasing resource efficiency, with some potential for achieving both. For example, if a dataset does not contain RFI that needs to be removed then no flagging action is taken, saving time and computing resources. If a dataset contains multiple levels of RFI, the system is capable of deciding whether or not to flag multiple times to achieve a better image. This also has potential to reduce human overhead in the entire end-to-end processing for data (calibration, imaging, quality assurance, automated products, reprocessing, end-user fine tuning, etc.).

\subsection{Automated Heuristic Discovery}\label{sec:p5.1.2}
The boundaries of the action regions can be thought of as heuristics. Traditionally, heuristic search is something that \textit{teams} of scientists and developers spend time researching and testing for traditional pipelines. Through simulated data and experimenting with different actions these regions and their boundaries are discovered. This moves the time-intensive task for humans to the machine. In our tests, we showed the decision regions are not only found but were clearly separable, quantifiable, interpretable, and aligned with domain expertise. We demonstrated RL is capable of handling single decisions and sequences of decisions in a non-greedy way. The sequences of decisions can be expanded to include any size decision chain. We intentionally chose a simplified environment to test this but it served as a proof of concept to warrant application to more complex data and decision scenarios.

\subsection{A Diagnostic to Evaluate Software Tools}\label{sec:p5.1.3}
Measuring how well a given processing action (tool) does on a given datasets gives us information on the performance of the tool. This measured quantity can give us feedback for how well the tool is achieving its purpose. This can be insightful for investigating a tool's performance or efficiency in certain situations.

\subsection{Moving Towards Real Data}\label{sec:p5.2}
The scope of this research was intentionally limited to a simplified subset of a typical data reduction workflow and environment to provide a proof of concept with unambiguous validation. This impacted the complexity of the datasets, the number of features used to estimate the state, the actions available to choose from, the amount of parameter tuning for each action, and the number of steps needed to get to an optimal state. Each of these constraints will need to be relaxed as we take this research forward to real data. 

In addition, the size of the datasets was kept small along with the number of actions and steps needed for multiple reasons. Not only was this done for clear interpretation but this kept the computing requirements to a reasonable size. With only two action steps and 6 available actions to choose from, each of the ~4000 datasets needed to be processed 36 different ways. We brute-forced every single pathway to have the ground truth to validate our results. As we move towards real data, more action choices and longer sequences of steps will be needed to process the data. As such, brute-forcing every possible pathway will become less reasonable. We will need to take advantage of more sophisticated RL methods beyond Q-learning.

With more complex data and longer processing sequences there is likely to be more complex correlations in high-dimensional space. The level of interpretability demonstrated here will become more challenging to extract but this provides an area for doing cutting-edge research.

\newpage
\section{Future Work}\label{sec:p5.3}
We have built this prototype to demonstrate the basics of applying RL to this problem and validate the results upon which to build more complex scenarios that solve real world problems. To that end we are expanding the scope of the environment as well as the RL methods used. For the environment, we will increase the number of input features, actions available, and stages of processing. Most notably will be the inclusion of the imaging stage. This stage will have a few different deconvolution algorithms to pick from. With the inclusion of the imaging stage and a few default deconvolution algorithm choices (and the simulations to warrant them), we anticipate RL methods will be able to demonstrate ‘self-calibration’ functionality \citep{1989ASPC....6..185C}. Additionally, we will build on the foundational RL methods used here towards more complex Reinforcement Learning methods such as Deep-Q-Networks and its derivatives, which combine neural networks with Q-learning. Increasing the scope and complexity of the problems needing solving will undoubtedly require additional computing requirements and therefore require thoughtful ways to keep training expenses manageable. While the path ahead is uncertain and difficult we believe the end result will far outweigh the cost.

\section{Acknowledgements}
Support for this work was provided by the NSF through the Grote Reber Fellowship Program administered by Associated Universities, Inc./National Radio Astronomy Observatory. We acknowledge the use of the CASA software package and NRAO computing resources. We thank the anonymous reviewer for their thoughtful consideration of our work and for pointing us to information about related efforts. We would like to thank Enrique Cesar Garcia from ESO for discussions about the relation between our work and the EDPS system. The National Radio Astronomy Observatory is a facility of the National Science Foundation operated under cooperative agreement by Associated Universities, Inc.

\clearpage
\bibliography{main}{}
\bibliographystyle{aasjournal}

\end{document}